\documentclass[twocolumn,aps,pra,superscriptaddress,showpacs,tightenlines,floatfix]{revtex4-2}
\usepackage{setspace}
\usepackage[T1]{fontenc}
\usepackage[latin9]{inputenc}
\usepackage{amstext}
\usepackage{amsmath}
\usepackage{amsfonts}
\usepackage{graphicx}
\usepackage{subfigure}
\usepackage{esint}
\usepackage{color}
\usepackage[colorlinks, linkcolor=blue, citecolor=blue,hyperindex,bookmarks=false,pdfstartview=FitH]{hyperref}
\usepackage{lineno}
\usepackage{soul}
\def\bibsection{\section*{REFERENCE}} 
\makeatletter

\@ifundefined{textcolor}{}
{%
 \definecolor{BLACK}{gray}{0}
\definecolor{WHITE}{gray}{1}
 \definecolor{RED}{rgb}{1,0,0}
 \definecolor{GREEN}{rgb}{0,1,0}
 \definecolor{BLUE}{rgb}{0,0,1}
 \definecolor{CYAN}{cmyk}{1,0,0,0}
 \definecolor{MAGENTA}{cmyk}{0,1,0,0}
 \definecolor{YELLOW}{cmyk}{0,0,1,0}
}

\makeatother

\begin{document}

\title{State Preparation in a Jaynes-Cummings Lattice with Quantum Optimal Control}

\author{Prabin Parajuli}
\affiliation{School of Natural Sciences, University of California, Merced, California 95343, USA}
\author{Anuvetha Govindarajan}
\affiliation{School of Natural Sciences, University of California, Merced, California 95343, USA}
\author{Lin Tian}
\email{ltian@ucmerced.edu}
\affiliation{School of Natural Sciences, University of California, Merced, California 95343, USA}

\begin{abstract}
High-fidelity preparation of quantum states in an interacting many-body system is often hindered by the lack of knowledge of such states and by limited decoherence times. Here we study a quantum optimal control (QOC) approach for fast generation of quantum ground states in a finite-sized Jaynes-Cummings lattice with unit filling. Our result shows that the QOC approach can generate quantum many-body states with high fidelity when the evolution time is above a threshold time, and it can significantly outperform the adiabatic approach. We study the dependence of the threshold time on the parameter constraints and the connection of the threshold time with the quantum speed limit. We also show that the QOC approach can be robust against control errors. Our result can lead to advances in the application of the QOC for many-body state preparation.
\end{abstract}
\maketitle

{\parindent 0 pt \bf INTRODUCTION} 
\vskip 2mm

{\parindent 0 pt Recent progresses in manipulating quantum states and dynamics in noisy intermediate-scale quantum (NISQ) devices have demonstrated the potential to solve complicated problems with various platforms~\cite{Preskill_2018, googlesupremacy_2019, photonicsupremacy_2020}. 
An important question among such problems is the preparation of many-body states with high fidelity using NISQ devices, which is crucial for quantum simulation, quantum metrology and quantum communication~\cite{QSimuReview1,QSimuReview2, QSensingReview,QCommunReview}. 
In the past, a number of approaches have been developed to generate desired quantum many-body states, including adiabatic processes~\cite{FarhiScience2001, Albash2016}, quantum shortcut approach~\cite{review_shortcut2019}, quantum phase estimation ~\cite{Kitaev1995, Abrams1997}, quantum eigensolvers~\cite{Peruzzo2014, Dumitrescu2018, LongResearch2020}, and open system approach~\cite{Kraus2008, Verstraete2009}. However, due to the intrinsic complexity of quantum many-body systems, it remains challenging to prepare such states with high accuracy.}

With quantum control techniques, precisely engineered pulse sequences have been employed to manipulate quantum states with high accuracy~\cite{QCBook2008}.
Among such techniques, the quantum optimal control (QOC) approach~\cite{KrotovBook, Rabitz_88, Werschnik_2007, Magann_2021} provides a computational framework to generate desired quantum states or quantum processes by searching for optimal, time-dependent control parameters under given constraints. 
In recent years, the QOC has been widely used in a broad range of applications from the implementation of high-fidelity quantum logic gates, the suppression of environmental noise, the control of quantum transduction processes, the generation of novel entangled states, to the control of quantum many-body systems~\cite{KhanejaJMR2005, QOCApplication, Marsden_2021,Bharti_2021, DoriaPRL2011}. 
The problem of preparing quantum states or processes can be formulated into an optimization problem in the QOC, where an algorithm is adopted to minimize the cost function.

Here we study the QOC approach for the preparation of many-body states in a finite-sized Jaynes-Cummings (JC) lattice. In the thermodynamic limit, a JC lattice with integer fillings (i.e., the average number of excitations per lattice site is an integer) can exhibit a quantum phase transition between the Mott-insulating (MI) and superfluid (SF) phases~\cite{Hartmann:2006, Greentree:2006, Angelakis:2007, 2007RossiniPRL_JC, 2008NeilPra_BH, Noh2017Review}. 
At a finite size, the ground states of a JC lattice in the MI and SF regimes still exhibit distinctive behaviors~\cite{TianPRL2011, Seo2015:1, 2015TianScienceChina_QS}.
The preparation of the ground states in a JC lattice is non-trivial, especially in the intermediate range between the deep MI and deep SF phases. In \cite{KCaiNpj2021}, we employed an optimized adiabatic approach for the state preparation in a JC lattice.
In this work with the QOC approach, we adopt the chopped random basis (CRAB) algorithm~\cite{CanevaPRA2011, MullerRepProgPhys2021} to parametrize the time-dependent couplings of the JC lattice and use the Nelder-Mead approach to optimize these couplings. 
Our numerical result shows that when the total evolution time is above a threshold time $T_{\rm th}$, the QOC approach can generate the target state with a high fidelity above a designated threshold value, and it can significantly outperform the adiabatic approach. 
We find that the threshold time decreases and the average energy fluctuation increases with the constraints on the time-dependent couplings, which indicates the connection between the threshold time and the quantum speed limit (QSL)~\cite{QSL1, QSL2, QSL3, CanevaPRL2009, BukovPRX2019, JonesPRA2010}. 
Furthermore, our numerical simulation shows that the QOC approach can be robust against control errors in the time-dependent couplings.  
JC lattices have been explored theoretically and implemented experimentally in various systems, including circuit QED system, nanophotonic devices, atoms, and trapped ions~\cite{2009KochPra_QS, 2012HouckNP_JCQS, Xue2017, Hoffman:2011, KeelingPRL2012, HouckPRX2017, Sala2015Nanophotonics, Leper2011Atom, Ivanov2009Ion, Toyoda2013Ion, Debnath2018Ion, BWLi2022Ion}. This work can shed light on the application of the QOC in many-body state preparation and lead to deeper understanding of the QSL in preparing quantum many-body states. 
\vskip 4mm

{\parindent 0 pt \bf RESULTS} 
\vskip 2mm

{\parindent 0 pt \bf 1. JC Lattice}

{\parindent 0 pt A JC lattice is illustrated in Fig.~\ref{fig1}(a), where each unit cell of the lattice contains a two-level system (qubit) coupled to a cavity mode with coupling strength $g$, and adjacent cavities are coupled by photon hopping with hopping rate $J$. The Hamiltonian of this lattice can be written as $H_t =H_0 +H_{\rm int}$ ($\hbar = 1$). Here
\begin{figure}
    \includegraphics[clip, width=\linewidth]{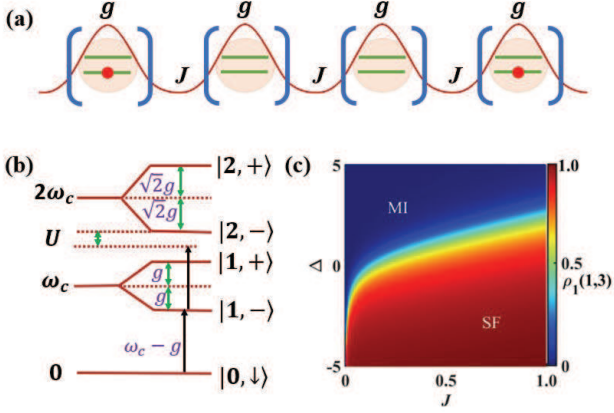} 
    \caption{{\bf Jaynes-Cummings Lattice. a} The schematic of a one-dimensional JC lattice with qubit-cavity coupling $g$ and photon hopping rate $J$.
    {\bf b} The energy spectrum of a single JC model for detuning $\Delta=0$, with $\vert 0,\downarrow\rangle$ the ground state and $\vert n,\pm\rangle$ ($n>1$, integer) the lowest excited states. 
    {\bf c} The single particle density matrix $\rho_{1}(1,3)$ vs hopping rate $J$ and detuning $\Delta$ for a $N=4$ lattice at unit filling. Here $g=1$, and all parameters are in dimensionless units~\cite{parameter_unit}.}
\label{fig1} 
\end{figure}
\begin{equation}
    H_0=\sum_{j=1}^N\left[\omega_c a_j^\dag a_j +\omega_{z}\frac{\sigma_{jz}+1}{2} +
    g\left (a_{j}^{\dagger}\sigma_{j-}+\sigma_{j+}a_j\right)\right]
    \label{eq:H0}
\end{equation}
is the Hamiltonian of the JC models in a finite-sized lattice of size $N$, with $j\in [1, N]$, $\omega_c$ the cavity frequency, $a_j$ ($a_j^\dagger$) the annihilation (creation) operator of the cavity modes, $\omega_{z}$ the energy splitting of the qubits, and $\sigma_{jz}, \sigma_{j\pm}$ the Pauli operators of the qubits. 
Also
\begin{equation}
    H_{\rm int}=-J\sum_{j=1}^N\left(a_{j}^{\dagger}a_{j+1}+a_{j+1}^{\dagger}a_{j}\right)
    \label{eq:Hint}
\end{equation}
describes photon hopping between neighboring sites in the lattice. We choose the periodic boundary condition with $a_{N+1}=a_1$ and denote $\Delta=\omega_{c}-\omega_{z}$ as the detuning between cavity and qubit frequencies.}

The qubit-cavity coupling $g$ induces a built-in nonlinearity in the energy spectrum of a single JC model~\cite{Larson2022Review}, which can be viewed as an effective onsite interaction with strength $U$, as shown in Fig.~\ref{fig1}(b). Details of the JC model spectrum can be found in the Supplementary Information. In the thermodynamic limit with $N\rightarrow\infty$, and at integer fillings when the number of excitations is an integer multiple of $N$, the competition between this onsite interaction and the photon hopping can lead to a quantum phase transition between the MI and SF phases ~\cite{Hartmann:2006, Greentree:2006, Angelakis:2007, 2007RossiniPRL_JC}. 
When dominated by the qubit-cavity coupling with $g\gg J$, the ground state of the JC lattice will be in a MI phase characterized by localized polariton excitations. In the limiting case of $J=0$, the ground state with $N$ excitations is the product state $\vert G\rangle_{J=0}=\prod_{j=1}^{N} \vert 1,-\rangle_j$ with each JC model in its first excited state $\vert 1,-\rangle_j$. 
When dominated by photon hopping with $J\gg g$, the ground state of the lattice will be in a SF phase with long range correlation. In the limiting case of $g=0$, the ground state is the Fock state $\vert G\rangle_{g=0} = \frac{1}{\sqrt{N!}} (a_{k=0}^\dagger)^N \vert 0,\downarrow\rangle$ with all excitations occupying the momentum-space mode $a_{k=0}=\frac{1}{\sqrt{N}}\sum_{j=1}^{N} a_j e^{i k\cdot j} $ for the quasi-momentum $k=0$. For a finite-sized lattice, the ground states also exhibit features of these phases in the corresponding parameter regimes~\cite{Seo2015:1, KCaiNpj2021}. These features can be illustrated with the single-particle density matrix $\rho_{1} (i,j)=\langle G|a_{i}^{\dagger}a_{j}|G\rangle/\langle G|a_{i}^{ \dagger}a_{i}|G\rangle$, which describes the spatial correlation between the cavity modes at sites $i$ and $j$, with $\vert G\rangle$ the ground state for given parameters. As shown in Fig.~\ref{fig1}(c), $\rho_{1} (1,3)$ for a $N=4$ lattice, and hence, the spatial correlation of the ground state, decrease algebraically (exponentially) in the SF (MI) phase. 
\vskip 2mm

{\parindent 0 pt \bf 2. Couplings and Fidelity}

{\parindent 0 pt Preparing the ground states of a JC lattice with integer fillings is a challenging task except for the limiting cases of $g=0$ or $J=0$. Here we will employ the QOC technique to achieve fast and high-fidelity state preparation in a JC lattice and compare our result with that of the adiabatic approach in \cite{KCaiNpj2021}. We define the fidelity of the prepared state with regards to the desired many-body ground state $\vert\psi_{\rm T}\rangle$ for the target parameters as 
\begin{equation}
	\mathbb{F}= \vert\langle\psi(T)|\psi_{\rm T}\rangle\vert^{\rm 2}, \label{eq:F}
\end{equation}
where $\vert\psi(T)\rangle$ is the state at the final time $T$ of the evolution. The cost function in the QOC is chosen as the infidelity $\mathbb{I}=1-\mathbb{F}$. 
The QOC approach minimizes the cost function by optimizing the coupling constants $g\left (t \right )$ and $J\left(t\right)$ in the Hamiltonian $H_t$~\cite{KrotovBook, Rabitz_88, Werschnik_2007, Magann_2021}. For simplicity of discussion, we let the detuning $\Delta(t)\equiv 0$ during the entire evolution. The couplings are bounded by the constraints $g_{\rm max}$ and $J_{\rm max}$, with $\vert g(t)\vert \le g_{\rm max}$ and $\vert J(t)\vert \le J_{\rm max}$ at an arbitrary time $t$. 
The numerical simulation is conducted on a JC lattice with four sites and four polariton excitations (i.e., unit filling). The initial Hamiltonian parameters are $g(0)=0$ and $J(0)=0.5$, and the target parameters are $g(T)=1$ and $J(T)=0.02$. The initial state of this  system is the ground state for the initial parameters, which is the SF state $\vert G\rangle_{g=0}$. The target state is the ground state for the target parameters, which is a MI state. 
During the evolution, the system is governed by the Hamiltonian $H_t $ with time-dependent couplings $g(t)$ and $J(t)$. 
We adopt the CRAB algorithm that parametrizes the couplings with truncated Fourier series~\cite{CanevaPRA2011, MullerRepProgPhys2021}, and apply the Nelder-Mead method for the optimization.} 

\begin{figure}
    \includegraphics[clip, width=\linewidth]{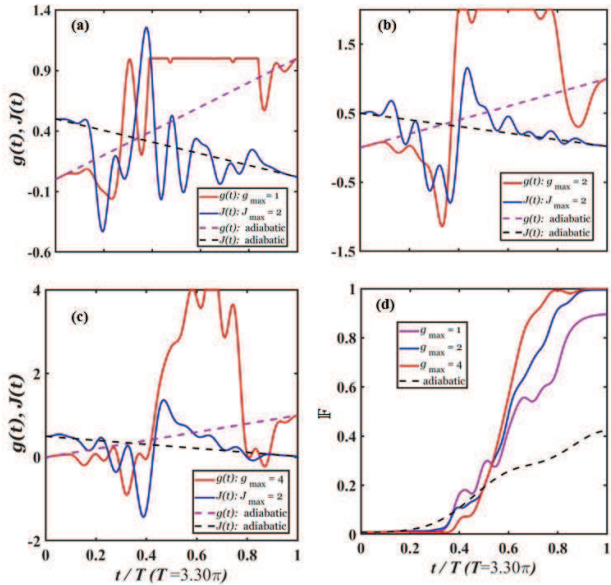}
    \caption{{\bf Optimized couplings. a-c} The optimized couplings $g(t)$ and $J(t)$ vs the relative evolution time $t/T$ for the constraints $J_{\rm max}=2$ and $g_{\rm max}=1, 2, 4$, respectively, with the total evolution time $T=3.30\pi$. {\bf d} The fidelity $\mathbb{F}$ vs $t/T$ for the couplings in {\bf a-c}. The dashed curves are for the adiabatic ramping.} 
    \label{fig2}
\end{figure}

In Fig.~\ref{fig2}(a-c), we plot the optimized couplings $g(t)$ and $J(t)$ vs the relative evolution time $t/T$ under the constraints $J_{\rm max}=2$ and $g_{\rm max}=1, 2, 4$, respectively, with total evolution time $T=3.30\pi$. The couplings in the adiabatic approach governed by (\ref{eq:g0}) and (\ref{eq:J0}) are plotted as dashed curves. The optimized couplings are continuous curves that change smoothly over the course of the evolution. For the constraint $g_{\rm max}=1$, $g(t)$ includes a large plateau at the maximal strength $g(t)=1$; whereas the plateau area decreases significantly for $g_{\rm max}=4$. In contrast, $J(t)$ has no plateau. This is because the system can already reach the deep SF phase when $J=J_{\rm max}=1$, and it does not require a larger value of $J$ to explore the Hilbert space. 
Our numerical result also shows that the fidelity for larger $g_{\rm max}$ is significantly higher than that for smaller $g_{\rm max}$, as shown in Fig.~\ref{fig2}(d). For $g_{\rm max}=2$ and $4$, the fidelity of the state at the final time exceeds the designated threshold fidelity $\mathbb{F}_{\rm th}=0.99$; while for $g_{\rm max}=1$, the fidelity cannot reach $0.99$. For all three $g_{\rm max}$, the fidelity at time $T$ is much higher than that from the adiabatic approach, demonstrating that the QOC approach can greatly outperform the adiabatic approach. 

\begin{figure}
    \includegraphics[clip, width=\linewidth]{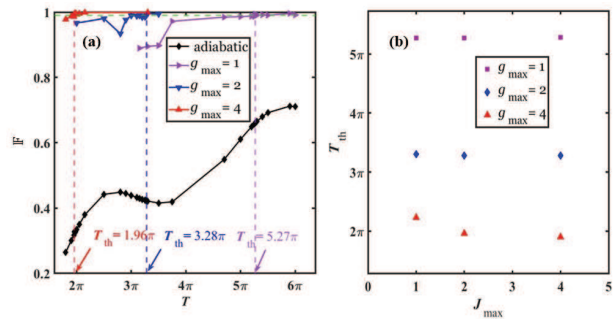}
    \caption{{\bf Fidelity and threshold time. a} The fidelity $\mathbb{F}$ of the prepared state vs the total evolution time $T$. The vertical dashed lines indicate the position of the threshold time $T_{\rm th}$ for each set of constraints. The constraints are $J_{\rm max}=2$ and $g_{\rm max}=1, 2, 4$. The dashed horizontal line corresponds to the threshold fidelity $\mathbb{F}_{\rm th}=0.99$.
    {\bf b} The threshold time $T_{\rm th}$ vs the constraint $J_{\rm max}$ for $g_{\rm max}=1, 2, 4$.}
    \label{fig3}
\end{figure}

The fidelity of the prepared states depends on the constraints $g_{\rm max}$, $J_{\rm max}$, and the total evolution time $T$.  
In Fig.~\ref{fig3}(a), we plot the fidelity vs $T$ for the constraints $g_{\rm max}=1, 2, 4$ and $J_{\rm max}=2$. The result shows that the fidelity exhibits an increasing trend with the total time $T$ and the constraint $g_{\rm max}$. Meanwhile, the fidelity from the QOC approach is significantly higher than the fidelity from the adiabatic ramping. For example, the QOC fidelity is greater than $0.99$ for $g_{\rm max}=2$, $J_{\rm max}=2$ and $T= 3.30\pi$, while the fidelity from the adiabatic ramping is only $0.42$ for the same evolution time $T$. 
\vskip 2mm

{\parindent 0 pt \bf 3. Threshold Time}

{\parindent 0 pt In the numerical simulation, we observe that when the total evolution time $T$ is below a threshold time $T_{\rm th}$, the QOC process cannot achieve a fidelity that is higher than the designated threshold fidelity, which we choose to be $\mathbb{F}_{\rm th}=0.99$.  
In Fig.~\ref{fig3}(a), the threshold time $T_{\rm th}$ for each set of constraints is indicated by a dashed vertical line. Our result shows that the threshold time decreases as the constraint $g_{\rm max}$ increases. Hence it will take less time to reach a desired fidelity when the coupling $g(t)$ can have a larger magnitude. 
To analyze the dependence of the threshold time on the constraints, we plot $T_{\rm th}$ vs the constraint $J_{\rm max}$ for different values of $g_{\rm max}$ in Fig.~\ref{fig3}(b). It is shown that $T_{\rm th}$ decreases significantly as $g_{\rm max}$ increases, but only decreases slightly when $J_{\rm max}$ increases. 
For the values of $g_{\rm max}$ used in our simulation, $J=J_{\rm max}=1$ is sufficiently large for the system to enter the deep SF phase. Thus the system does not demand larger value of $J$ or subsequently longer evolution time in order to reach high fidelity, which leads to $T_{\rm th}$'s weak dependence on $J_{\rm max}$. This result agrees with that of Fig.~\ref{fig2}(a-c), where $J(t)$ does not exhibit any plateau during the evolution. 
The threshold time for different constraints is given in Table.~\ref{tab1}, together with the fidelity for the total evolution time $T=T_{\rm th}$ from the QOC approach and from adiabatic ramping.}
\begin{table}[t]
    \caption{The threshold time $T_{\rm th}$ for selected constraints and the corresponding fidelity $\mathbb{F}$ for total evolution time $T=T_{\rm th}$ using QOC and using adiabatic ramping (adia.).}
    \label{tab1}
    \begin{tabular}{lccc} \hline\hline
			Constraints &
			$T_{\rm th}$ & 
			$\mathbb{F}$ (QOC) &
			$\mathbb{F}$ (adia.) \\ \hline
			$J_{\rm max}=1, g_{\rm max}=1$ & \hspace*{5mm}$5.27\pi$\hspace*{5mm} & \hspace{2mm}          0.9944 \hspace*{2mm} & 0.6610\\
			$J_{\rm max}=1, g_{\rm max}=2$ & $3.30\pi$ & 0.9932 & 0.4213\\
			$J_{\rm max}=1, g_{\rm max}=4$ & $2.23\pi$ & 0.9963 & 0.3995\\
			$J_{\rm max}=2, g_{\rm max}=1$ & $5.27\pi$ & 0.9944 & 0.6610\\
			$J_{\rm max}=2, g_{\rm max}=2$ & $3.28\pi$ & 0.9927 & 0.4223\\
			$J_{\rm max}=2, g_{\rm max}=4$ & $1.96\pi$ & 0.9954 & 0.3276\\
			$J_{\rm max}=4, g_{\rm max}=1$ & $5.28\pi$ & 0.9925 & 0.6626\\
			$J_{\rm max}=4, g_{\rm max}=2$ & $3.28\pi$ & 0.9927 & 0.4223\\
			$J_{\rm max}=4, g_{\rm max}=4$ & $1.90\pi$ & 0.9904 & 0.3001\\ \hline\hline
    \end{tabular}
\end{table}

We compare the threshold time $T_{\rm th}$ from our numerical simulation with an estimation of the quantum speed limit (QSL) $T_{\rm QSL}$, which is the minimal time for a given quantum system to evolve from an initial state to a target state~\cite{QSL1, QSL2, QSL3, CanevaPRL2009, BukovPRX2019, JonesPRA2010}. 
We estimate the QSL with~\cite{CanevaPRL2009}: 
\begin{equation}
    T_{\rm QSL}\approx \frac{\arccos{\left\vert\langle\psi(0)\vert\psi_{\rm T}\rangle\right\vert}}{\Delta E_{\rm ave}}. \label{eq:TQSL}
\end{equation}
Here, $\arccos{\left\vert\langle\psi(0)\vert\psi_{\rm T}\rangle\right\vert}$ describes the distance between the initial and the target states in the Hilbert space~\cite{JonesPRA2010}. For orthogonal states with $\langle\psi(0)\vert\psi_{\rm T}\rangle=0$, the distance is $\pi/2$. For the initial and target states in our simulation, the distance is $0.469 \pi$. 
Also $\Delta E_{\rm ave} = \frac{1}{T} \int_0^T dt \Delta E (t)$ is the average energy fluctuation during the time evolution with $\Delta E (t)=\sqrt{\langle [H_t - \langle H_t\rangle]^2 \rangle} $ being the instantaneous energy fluctuation of the Hamiltonian $H_t$ at time $t$, and the operator average is taken on the instantaneous quantum state $\vert\psi(t)\rangle$. 
\begin{figure}
    \includegraphics[clip, width=\linewidth]{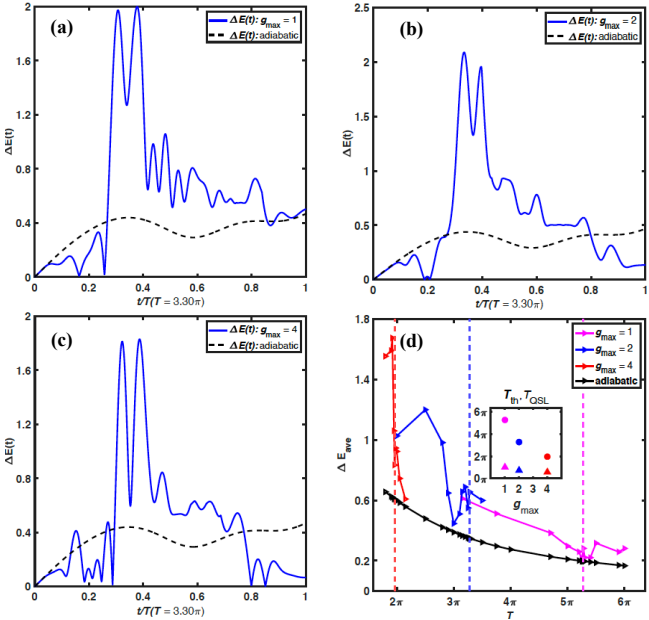}
    \caption{{\bf Energy fluctuation. a-c} The energy fluctuation $\Delta E(t)$ vs the relative evolution time $t/T$ for the constraints $J_{\rm max}=2$ and $g_{\rm max}=1, 2, 4$, respectively, and $T=3.30\pi$. The dashed curve is for the adiabatic ramping.
    {\bf d} The average energy fluctuation $\Delta E_{\rm ave}$ vs the total evolution time $T$ for the constraints in (a-c). The inset of (d) shows the threshold time $T_{\rm th}$ (circle) and the estimated $T_{\rm QSL}$ (triangle) vs $g_{\rm max}$.} 
    \label{fig4}
\end{figure}

In Fig.~\ref{fig4}(a-c), we plot the energy fluctuation $\Delta E(t)$ as a function of the relative evolution time $t/T$ for the same constraints $J_{\rm max}$, $g_{\rm max}$, and the same evolution time $T$ as those in Fig.~\ref{fig2}. The result from the adiabatic approach is plotted as the dashed curve. In all three plots, the energy fluctuation is the strongest when $t/T\in (0.3,\,0.4)$, and it is far stronger than the energy fluctuation in the adiabatic process. For $g_{\rm max}=2, 4$, $\Delta E(t)$ becomes very small when $t$ approaches the final time $T$, indicating that the final state occupies the ground state with high probability. For $g_{\rm max}=1$, $\Delta E(t)$ at $t=T$  remains large and is comparable to that from the adiabatic approach, which shows that the system has a sizable probability to be in the excited states in this case. 
This is because the threshold time for $g_{\rm max}=2, 4$ (for $g_{\rm max}=1$) is shorter (longer) than the evolution time $T=3.30\pi$, and hence the QOC process can (cannot) reach high fidelity. 
In Fig.~\ref{fig4}(d), we plot the average fluctuation energy $\Delta E_{\rm ave}$ vs the total evolution time $T$ for the constraints used in Fig.~\ref{fig4}(a-c). Here $\Delta E_{\rm ave}$ shows a decreasing trend as $T$ increases and is stronger than that from the adiabatic approach. 
Using (\ref{eq:TQSL}) and the result of $\Delta E_{\rm ave}$ for the threshold time $T_{\rm th}$, we estimate the QSL. As shown in the inset of Fig.~\ref{fig4}(d), the estimated $T_{\rm QSL}$ exhibits similar behavior to the threshold time $T_{\rm th}$, decreasing with the increase of the constraint $g_{\rm max}$. Meanwhile, the estimated QSL is comparable in scale to the threshold time, but it is shorter than the threshold time.
We note that this comparison is only qualitative. The estimation of the QSL presented here is a rough approximation due to the complexity of the JC lattice, and the threshold time is defined for a specific threshold fidelity chosen in our numerical simulation.
\vskip 4mm

{\parindent 0 pt \bf DISCUSSION} 
\vskip 2mm

{\parindent 0 pt \bf Control Error and Decoherence}
 
{\parindent 0 pt In superconducting quantum devices, tunable qubit-cavity coupling and cavity hopping (i.e., cavity coupling) can reach a few hundreds of MHz~\cite{squbit_rev, BlaisRMP2021cQED, SiddiqiPRL2021, CampbellLaHayePRApplied2023}. 
For example, tunable qubit-cavity coupling can be achieved via flux-tuned inductive coupling in the g-mon configuration or via a tunable coupler~\cite{Chen:2014, FYanPRApplied2018}. Tunable cavity hopping can be achieved by connecting cavities with a tunable Josephson junction~\cite{SandbergAPL2008}. 
We assume that the dimensionless coupling $g=1$ used in our numerical simulation corresponds to $g=2\pi\times 100$ MHz in actual devices~\cite{parameter_unit}. A dimensionless evolution time of $T=3.30\pi$ then corresponds to $T=16.5\,$ ns. The optimized, time-dependent couplings $g(t)$ and $J(t)$ need to be generated within this time scale, which can be implemented with current technology.} 

To explore the robustness of the QOC approach against control errors, we simulate the errors by adding a time-dependent Gaussian noise to the optimized solutions of $g(t)$ and $J(t)$ with
\begin{subequations}
    \begin{align}	
	g(t)& \rightarrow g(t)+\delta_{1}(t), \label{eq:dg}\\
	J(t)& \rightarrow J(t)+\delta_{2}(t), \label{eq:dJ} 
    \end{align}	
\end{subequations}
where $\delta_{1}(t)$ and $\delta_{2}(t)$ are Gaussian noise at time $t$ with the standard deviation $\sigma$. We obtain the fidelity of the prepared state in the presence of these errors. For a given value of  $\sigma$, we conduct the simulation on $1000$ samples of the time-dependent errors and calculate the average value of the fidelity.  The total evolution time is chosen to be the threshold time $T_{\rm th}$ for given constraints.
\begin{figure}
    \includegraphics[width=\linewidth]{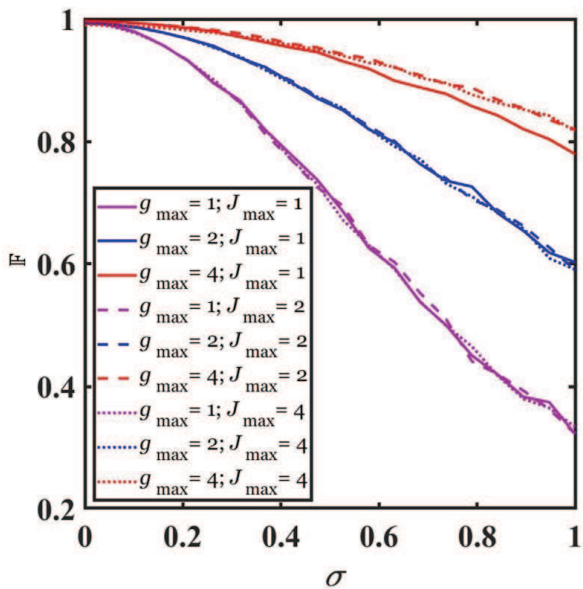}
    \caption{{\bf Effect of control error.} The fidelity vs the standard deviation $\sigma$ of Gaussian control errors. The total evolution time for given constraints is the corresponding threshold time $T_{\rm th}$ in Table~\ref{tab1}.}
    \label{fig5} 
\end{figure}
Fig.~\ref{fig5} shows that the fidelity of the prepared state decreases with the standard deviation of the control errors. We observe that the decrease of the fidelity for larger $g_{\rm max}$ is slower than that for smaller $g_{\rm max}$; whereas the decrease of the fidelity remains almost the same for different values of $J_{\rm max}$. For $J_{\rm max}=2$ and $\sigma=0.05$, the fidelity is reduced to $\mathbb{F}=0.9902$, $0.9910$, $0.9948$ for $g_{\rm max}=1$, $2$, $4$, respectively. Compared with the fidelity for no control errors given in Table~\ref{tab1}, the reduction of the fidelity is negligible. This result shows that the QOC approach can be robust against control error.

Another factor that could affect the fidelity of the QOC process is the decoherence of the qubits and the cavity modes. In the NISQ era, the finite decoherence time of the quantum devices sets a limit on the time for coherent evolution. In a JC lattice with unit filling, the polariton excitations can decay in a time scale comparable to the decoherence times. 
At the current state-of-the-art, the decoherence time of superconducting qubits can reach $\sim 100 {\rm \mu s}$~\cite{squbit_rev, BlaisRMP2021cQED, SiddiqiPRL2021}. Superconducting cavities can have quality factor greater than $10^6$. For a cavity frequency of $\omega_c=2\pi\times 5$ GHz, such a quality factor gives a cavity decay time of $\sim 32\mu$s. With an evolution time of $16.5$ ns, the QOC can be completed in a much shorter time than the decoherence time of superconducting qubits and cavities, and hence the effect of decoherence can be neglected. This analysis has been confirmed by our numerical simulation using a master equation approach, as detailed in the Supplementary Information.
We want to note that the short evolution time required in the QOC approach is one of its advantages over the adiabatic approach. 
\vskip 4mm

{\parindent 0 pt \bf METHODS} 
\vskip 2mm

{\parindent 0 pt In the CRAB algorithm used in our simulation, the time-dependent parameters $g(t)$ and $ J(t)$ are parametrized with truncated Fourier series to the $8$th harmonics and can be written as~\cite{CanevaPRA2011, MullerRepProgPhys2021}
\begin{subequations}
    \begin{align}
		 g(t)& =g_{0}(t)\left[ 1 +s(t)f_1(t)\right], \label{eq:gt}\\
		 J(t)& =J_{0}(t)\left[1 +s(t) f_2(t) \right], \label{eq:Jt}
    \end{align}	
\end{subequations}  
with 
\begin{subequations}
    \begin{align}
        f_1(t) &= \sum_{k=1}^{8} c_{1,k} \cos \left(\frac{\omega_{1,k}t}{T}\right) +c_{2,k} \sin \left(\frac{\omega_{1,k}t}{T}\right), \label{eq:f1t} \\
        f_2(t) &= \sum_{k=1}^{8} d_{1,k} \cos \left(\frac{\omega_{2,k}t}{T}\right)+d_{2,k} \sin \left(\frac{\omega_{2,k}t}{T}\right), \label{eq:f2t}
    \end{align}	
\end{subequations}  
where $c_{i,k}$ and $d_{i,k}$ ($i=1,\,2$ and $k\in[1,8]$) are the Fourier coefficients of the $k$-th harmonics in $g(t)$ and $J(t)$, respectively, and $\omega_{i,k}=k+\delta\omega_{i,k}$ is the frequency of the $k$-th harmonics with an adjustable offset $\delta\omega_{i,k}$. 
Here, $g_{0}(t)$ [$J_{0}(t)$] is the linear ramping function for the coupling $g$ ($J$) in the adiabatic approach with 
\begin{subequations}
    \begin{align}
	g_{0}(t)& =g(0)+ \left[g(T)-g(0)\right]t/T,\label{eq:g0}\\
		J_{0}(t)& =J(0)+\left[J(T)-J(0)\right]t/T, \label{eq:J0}
    \end{align}
\end{subequations}
where $g(0)$, $J(0)$ [$g(T)$, $J(T)$] are the initial (target) values for the couplings. 
The function $s(t)=\left[1-\cos(2\pi t/T)\right]$. With $s(0)=s(T)=0$, it ensures that the initial and final values of $g(t)$ [$J(t)$] are the same as that of $g_0(t)$ [$J_0(t)$]. During the QOC process, $g(t)$ is bounded by the constraint $g_{\rm max}$ with 
\begin{equation}
    g(t) = \begin{cases}
    g(t), & \textrm{if\quad} \vert g(t)\vert \le  g_{\rm max} \\
    g_{\rm max} \frac{g(t)}{\vert g(t)\vert}, & \textrm{if\quad} \vert g(t)\vert  > g_{\rm max};
    \end{cases}
\end{equation}
and similarly, $J(t)$ is bounded by the constraint $J_{\rm max}$.}

For a given set of initial (target) parameters for the JC lattice, we obtain the initial state $\vert\psi_0\rangle $ (the target state $\vert\psi_{\rm T}\rangle $) by diagonalizing the corresponding Hamiltonian $H_{t}$.
The optimization process begins with a random set of parameters $c_{i,k}$, $d_{i,k}$, and $\delta\omega_{i,k}$ and has a maximum of $150000$ iterations. The convergence of these parameters vs the iteration number $n$ can be found in Fig.~S1 of the Supplementary Information. 
\vskip 4mm

{\parindent 0 pt \bf DATA AVAILABILITY}
\vskip 2mm

{\parindent 0 pt The data that support the findings of this study are available from the corresponding author upon reasonable request.}
\vskip 4mm

{\parindent 0 pt \bf CODE AVAILABILITY}
\vskip 2mm

{\parindent 0 pt The codes that are used to produce the data presented in this study are available from the corresponding author upon reasonable request.}
\vskip 4mm

{\parindent 0 pt \bf ACKNOWLEDGEMENTS} 
\vskip 2mm

{\parindent 0 pt This work is supported by the NSF Award No. 2037987 and the UC-MRPI Program (Grant ID M23PL5936).}
\vskip 4mm

{\parindent 0 pt \bf AUTHOR CONTRIBUTIONS}
\vskip 2mm

{\parindent 0 pt P.P. developed the QOC algorithm in the numerical simulation, P.P. and A.G. conducted the simulation and generated the figures, L.T. designed the project and analyzed the numerical data, P.P. and L.T. wrote the paper, and all authors discussed the results and contributed to the final paper.}
\vskip 4mm

{\parindent 0 pt \bf COMPETING INTERESTS}
\vskip 2mm

{\parindent 0 pt The authors declare that there are no competing interests.}
\vskip 4mm

{\parindent 0 pt \bf ADDITIONAL INFORMATION}
\vskip 2mm

{\parindent 0 pt {\bf Supplementary Information.} The online version contains supplementary materials available at:}

{\widetext
\newpage
\setcounter{equation}{0}
\setcounter{figure}{0}

\def\bibsection{\section*{SUPPLEMENTARY REFERENCES}} 
\renewcommand{\theequation}{S\arabic{equation}}
\renewcommand{\thefigure}{S\arabic{figure}}
\renewcommand{\thetable}{S\arabic{table}}
\renewcommand{\bibnumfmt}[1]{[S#1]}
\renewcommand{\citenumfont}[1]{S#1}

{\parindent 0 pt \bf SUPPLEMENTARY NOTES}
\vskip 2mm

{\parindent 0 pt \bf Energy Spectrum of JC Model} 

{\parindent 0 pt We consider a single Jaynes-Cummings (JC) model on site $j\in [1, N]$ of the lattice with the Hamiltonian 
\begin{equation}
    H_{j0}=\left[\omega_c a_j^\dag a_j +\omega_{z}\frac{\sigma_{jz}+1}{2} +
    g\left (a_{j}^{\dagger}\sigma_{j-}+\sigma_{j+}a_j\right)\right]
    \label{Seq:Hj0}
\end{equation}
with $\omega_c$ the frequency of the cavity mode, $a_j$ ($a_j^\dagger$) the annihilation (creation) operator of the cavity mode, $\omega_{z}$ the energy splitting of the qubit, and $\sigma_{jz}, \sigma_{j\pm}$ the Pauli operators of the qubit. We describe the eigenstates of the JC model using the basis set $\{|n,s\rangle\}$, which are the product states of the photon number state $\vert n\rangle$ ($n$ being integer) of the cavity and the spin up or down state ($s=\uparrow$ or $\downarrow$) of the qubit.} 

The eigenstates of the JC model include the ground state $|g_{0}\rangle=|0,\downarrow\rangle$ with zero photon and the qubit in the spin down state (i.e., no excitation in the model), and the polariton doublets $|n,\pm\rangle$ with $n$ excitations:
\begin{subequations}
    \begin{align}
    |n,+\rangle & =  {\cos(\theta/2)}|n,\downarrow\rangle +{\sin(\theta/2)}|n-1,\uparrow\rangle,     \label{Seq:eigst1} \\
    |n,-\rangle &=  {\sin(\theta/2)}|n,\downarrow\rangle -{\cos(\theta/2)}|n-1,\uparrow\rangle,
    \label{Seq:eigst2}
    \end{align}
\end{subequations}
where $\theta={2\arcsin}\sqrt{[1-\Delta/\chi(n)]/2}$, $\chi(n)=\sqrt{\Delta^{2}+4ng^{2}}$, and $\Delta=\omega_{c}-\omega_{z}$ is the detuning between the cavity and the qubit~\cite{BlaisPRA2004}. 

The eigenenergies for different eigenstates are $E_{g_{0}}=0$, and $E_{n,\pm}=n\omega_{c}-\frac{1}{2}\Delta \pm \frac{\chi(n)}{2}$. When the coupling strength $g$ is nonzero, the energies of the eigenstates are not equally spaced with, e.g., $(E_{n+1,-}-E_{n,-})>(E_{n,-}-E_{n-1,-})$. This nonlinearity in the energy spectrum is at the root of many interesting phenomena in the JC model (and hence JC lattices), such as the photon blockade effect and the Mott insulator-superfluid phase transition.
\vskip 2mm

{\parindent 0 pt \bf QOC Algorithm and Optimization Parameters} 

{\parindent 0 pt In our numerical simulation, we first diagonalize the Hamiltonian of the JC lattice at the initial parameters $g(0)$, $J(0)$ and $\Delta(0)$ to find its ground state $\vert \psi_{0}\rangle$. This state is used as the initial state of the evolution with $|\psi(0)\rangle =\vert \psi_{0}\rangle$. Here, the initial parameters are chosen so that $\vert \psi_{0}\rangle $ is easy to prepare in realistic systems~\cite{KCaiNpj2021}. We then solve the ground state of the Hamiltonian at the target parameters $g(T)$, $J(T)$ and $\Delta(T)$, which is the desired many-body state $\vert \psi_{\rm T}\rangle $ to be prepared at the final time $T$ of the evolution.}

In our quantum optimal control (QOC) approach, we adopt the chopped random basis (CRAB) algorithm to parameterize the couplings $g(t)$ and $J(t)$ with truncated Fourier series. The parameters $c_{i,k}$, $d_{i,k}$ and $\delta\omega_{i,k}$ ($i=1,2$ and $k\in[1,8]$ being integers), as defined in Eq.~(7a) and (7b) in the Methods section of the main paper, include a total of $48$ optimization parameters.
In the beginning of the optimization process, a random set of $c_{i,k}$, $d_{i,k}$ and $\delta\omega_{i,k}$ are used to generate the initial trial functions for $g(t)$ and $J(t)$. Note that the choice of the random parameters only affects the values of $g(t)$ and $J(t)$ at $0<t<T$. At $t=0$ ($t=T$), the couplings always remain the same as the initial (target) parameters of the JC lattice. 
During the optimization process, following the protocol in  Eq.~(9) in the Methods section of the main paper, the values of $g(t)$ and $J(t)$ are bounded by the constraints $g_{\rm max}$ and $J_{\rm max}$, respectively. 
Given a set of time-dependent couplings $g(t)$ and $J(t)$, the system evolves from the initial state $\vert \psi_{0}\rangle$ under the Hamiltonian $H_t\left [g(t), J(t)\right ]$ to reach the final state $\vert \psi(T) \rangle$ at time $T$. The cost function for this set of couplings can  then be calculated. 
In this work, we use the Nelder-Mead method to minimize the cost function and find the optimal values for $c_{i,k}$, $d_{i,k}$ and $\delta\omega_{i,k}$. 

In our simulation, we choose the maximal number of iterations to be 150000. For a given total time $T$, if the optimization process converges to a fidelity above the threshold value $\mathbb{F}_{\rm th}=0.99$ within the maximal number of iterations, we consider the QOC process successful. The convergence of the QOC process is determined by the difference between the optimization parameters in adjacent iterations. If the difference is smaller than the pre-defined error tolerance, then we consider that convergence has been achieved. 
In Fig.~\ref{figS1}(a-f), we plot the Fourier coefficients $c_{i,k}$, $d_{i,k}$ and the frequency offsets $\delta\omega_{i,k}$ vs the iteration number $n$ under the constraints $J_{\rm max}=1$, $g_{\rm max}=2$ and for the total evolution time $T=T_{\rm th}=3.30\pi$. From the numerical result, we find that the number of iterations for this simulation to converge is $n=43194$. 
\begin{figure}
    \subfigure[The coefficient $c_{1,k}$ vs the iteration number $n$.]{
	\includegraphics[clip, width=8cm]{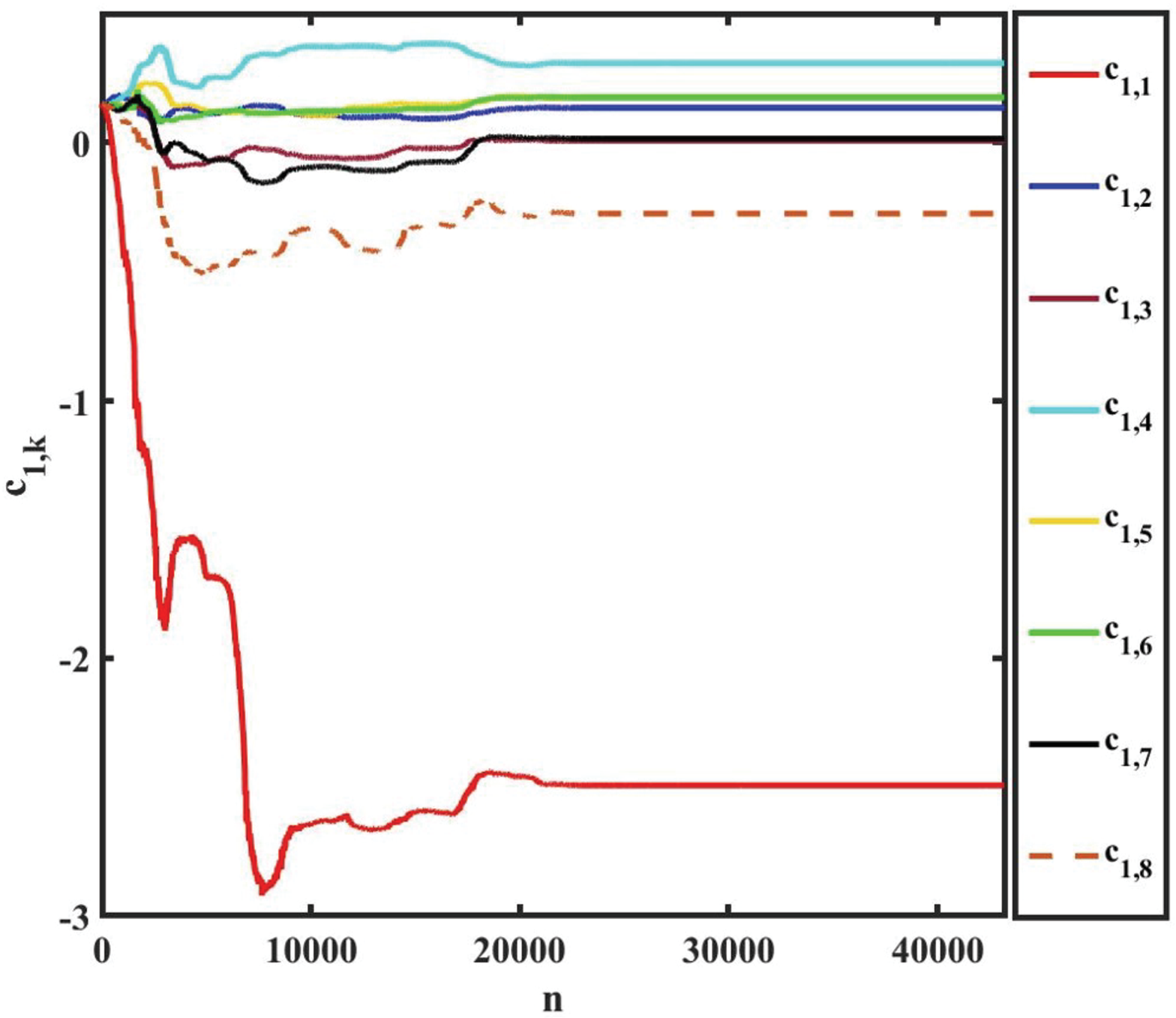}}
    \subfigure[The coefficient $c_{2,k}$ vs the iteration number $n$.]{
	\includegraphics[clip, width=8cm]{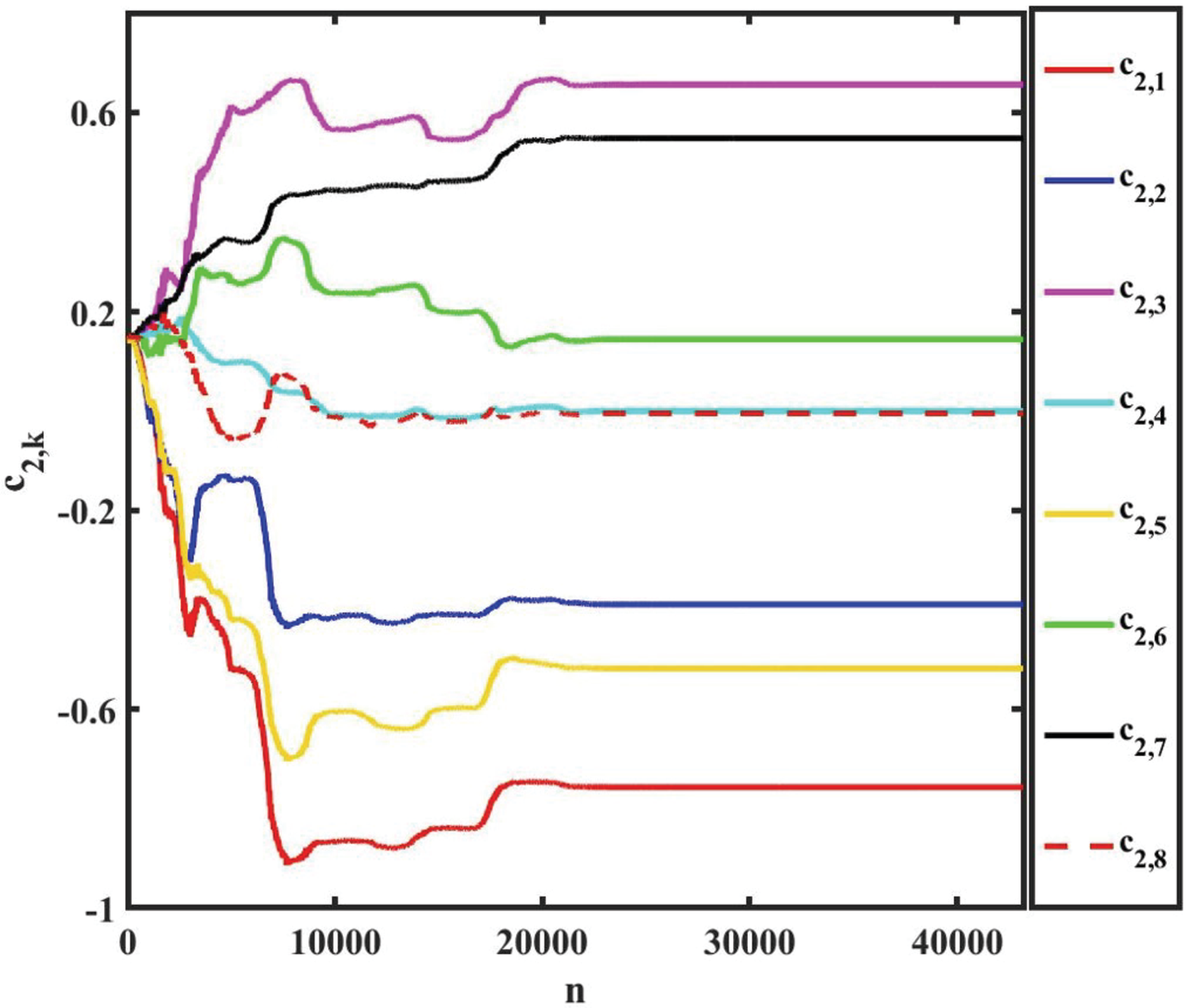}} \\ 
    \subfigure[The coefficient $d_{1,k}$ vs the iteration number $n$.]{
		\includegraphics[clip, width=8cm]{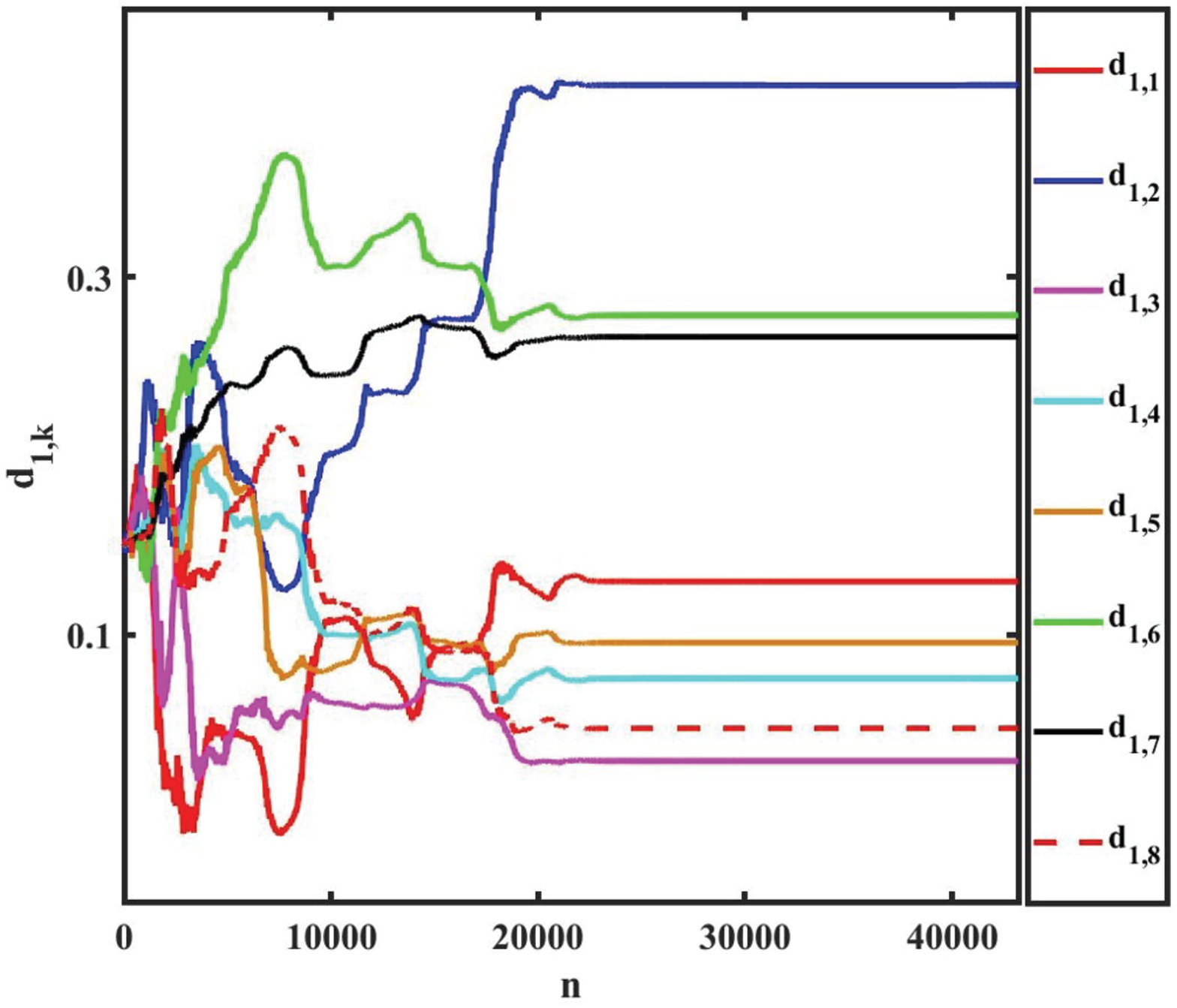}}
    \subfigure[The coefficient $d_{2,k}$ vs the iteration number $n$.]{
	\includegraphics[clip, width=8cm]{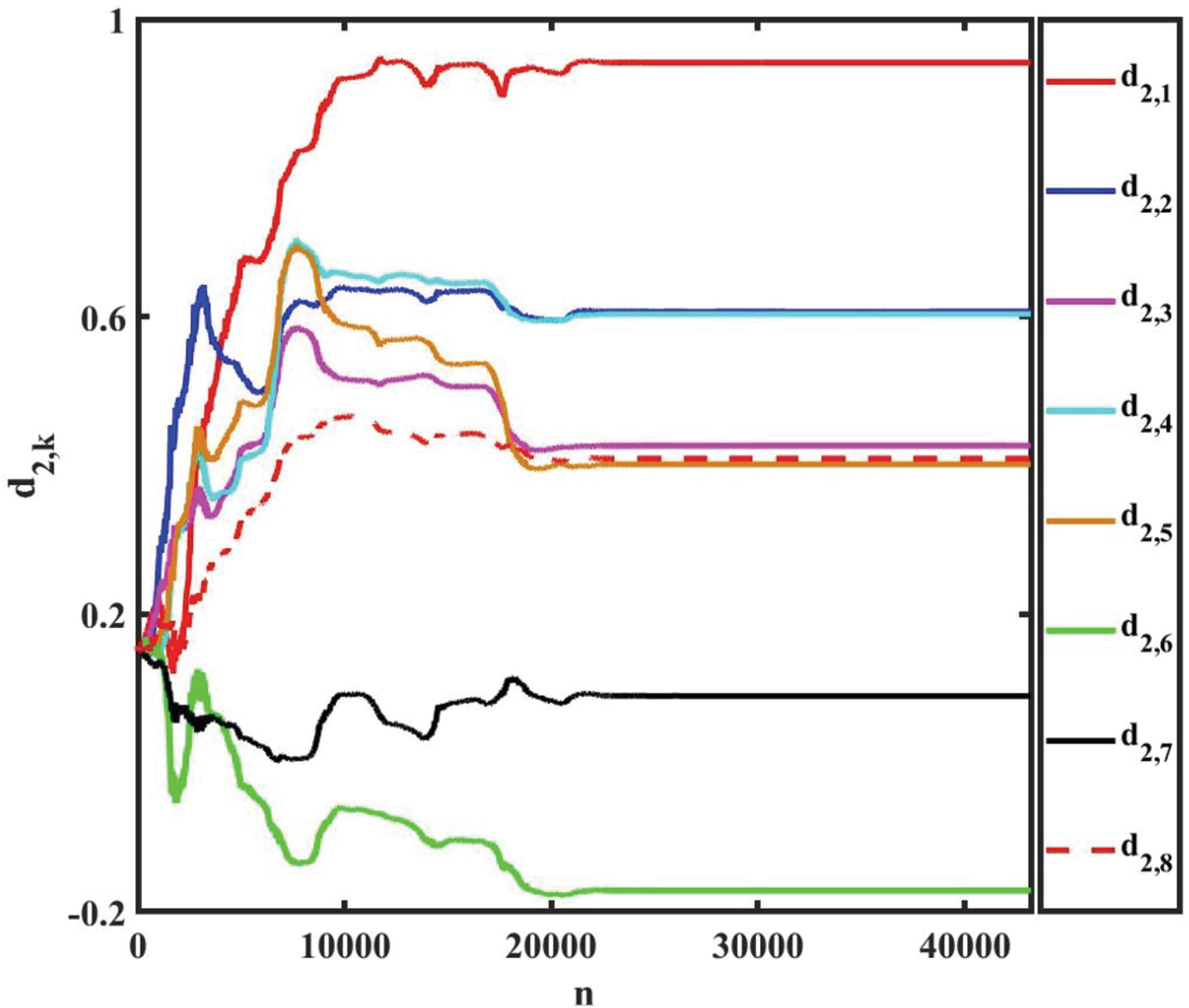}} \\
    \subfigure[The offset $\delta\omega_{1,k}$ vs the iteration number $n$.]{
		\includegraphics[clip, width=8cm]{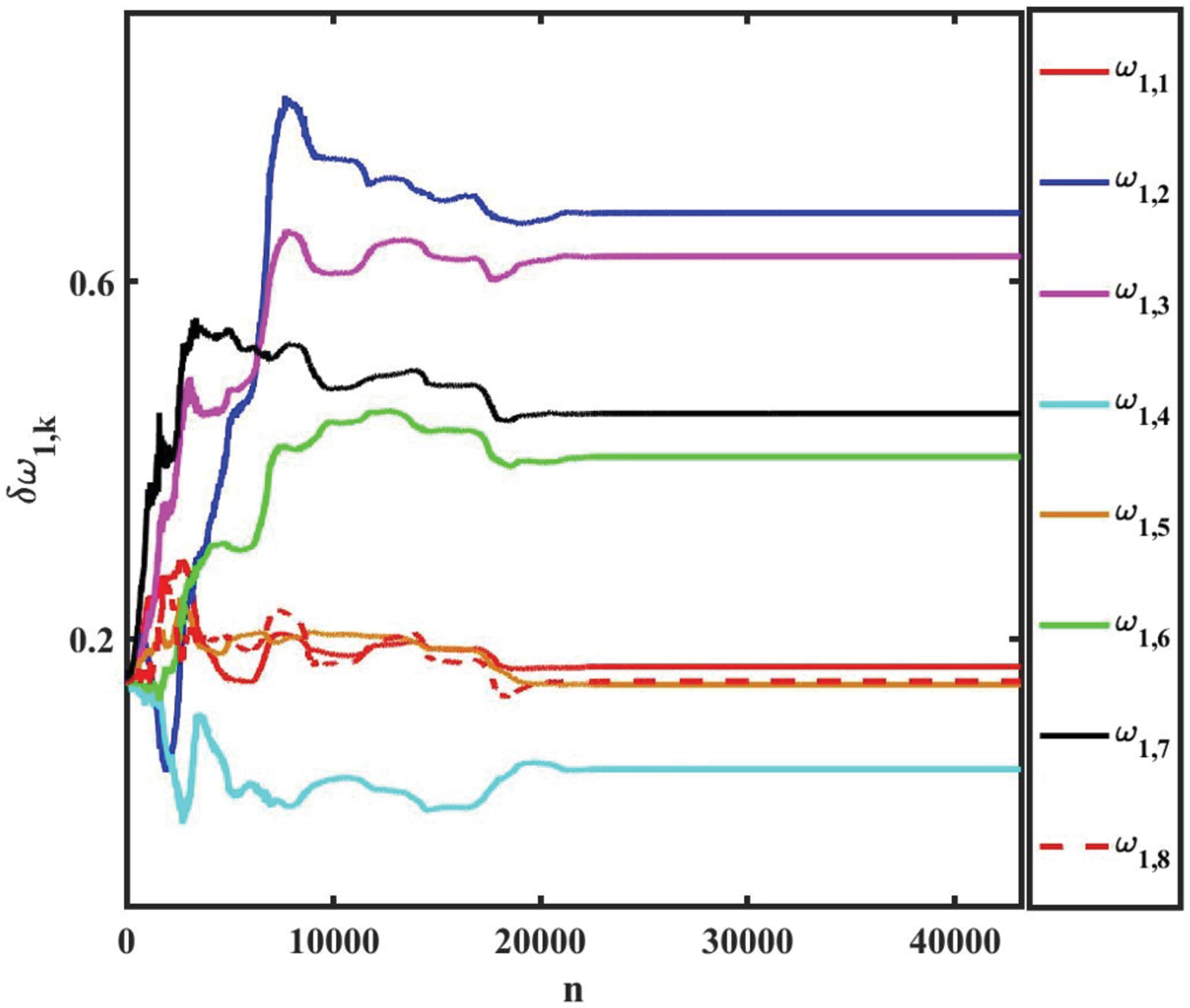}} 
    \subfigure[The offset $\delta\omega_{2,k}$ vs the iteration number $n$.]{
		\includegraphics[clip, width=8cm]{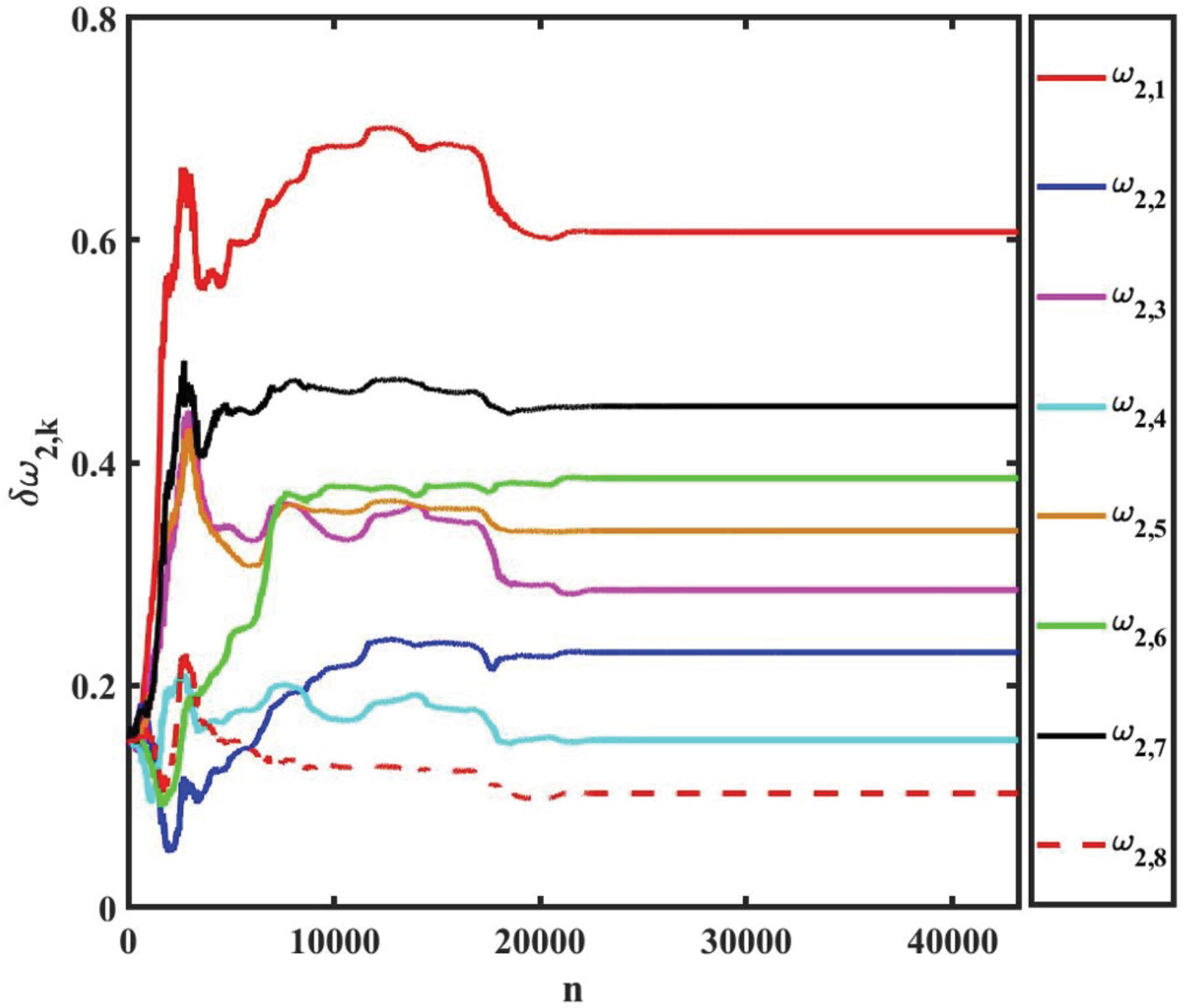}}
\caption{Optimization parameters $c_{i,k}$, $d_{i,k}$, and $\delta\omega_{i,k}$ vs the iteration number $n$. The constraints for the couplings are $J_{\rm max}=1$ and $g_{\rm max}=2$. The evolution time is  $T=T_{\rm th}=3.30\pi$.}
\label{figS1}
\end{figure}
\vskip 2mm

{\parindent 0 pt \bf Master Equation Approach for Decoherence}

{\parindent 0 pt To quantitatively characterize the effect of decoherence, we adopt the following master equation~\cite{WallsMilburnBook}: 
\begin{equation}
\frac{d\rho}{dt} = -i \left [ H_{t} , \rho \right] 
+  \frac{\kappa}{2} \sum_{j=1}^{N} \left(2 a_{j} \rho a_{j}^{\dag} - a_{j}^{\dag} a_{j} \rho -  \rho a_{j}^{\dag} a_{j} \right)  
+  \frac{\gamma}{2} \sum_{j=1}^{N} \left(2 \sigma_{j-} \rho \sigma_{j+} - \sigma_{j+} \sigma_{j-} \rho -  \rho  \sigma_{j+} \sigma_{j-} \right),\label{eq:Srho}
\end{equation}  
where $\rho$ is the density matrix of the JC lattice, $H_{t}$ is the total Hamiltonian with the time-dependent, optimized couplings $g(t)$ and $J(t)$, $\kappa$ is the cavity decay rate, and $\gamma$ is the qubit decoherence rate. For simplicity of discussion, we assume that all cavity modes have the same decay rate and all qubits have the same decoherence rate. For qubit decoherence, we omit the pure dephasing term that can have the form  $\frac{\gamma_{d}}{2} \sum_{j=1}^{N} \left(\sigma_{jz} \rho \sigma_{jz} -\rho \right)$ with dephasing rate $\gamma_{d}$.  It can be shown that the dephasing term will have comparable effect as the qubit and cavity decay terms in Eq.~(\ref{eq:Srho}). }

In the numerical simulation, we use the same parameters as discussed in the main paper. We take the energy unit to be $g  = 2\pi \times 100$ MHz, and give all parameters in dimensionless unit according to this energy unit. 
With a qubit decoherence time of $100 {\rm \mu s}$, the qubit decoherence rate can be derived as $\gamma  = 2\pi\times \frac{5}{\pi}$ kHz. 
In dimensionless unit, this corresponds to a decoherence rate $\gamma = \frac{5}{\pi} \times 10^{-5}$.  
With a cavity frequency of $\omega_{c} = 2\pi \times 5$ GHz and a quality factor of $Q=10^{6}$, the cavity decay rate is $\kappa=\omega_{c}/Q = 2\pi \times 5$ kHz. In dimensionless unit, this means $\kappa = 5\times 10^{-5}$. 
We run the master equation to test the effect of the decoherence terms with the three sets of parameters used in Fig.~4 of the main paper, i.e., $J_{\rm max}=2$, $g_{\rm max}=1,\,2,\,4$, respectively, and $T=3.3\pi$. 
The fidelities of the final state without and with the decoherence terms are given in Table.~\ref{tabS1}. 
\begin{table}[t]
    \caption{The fidelity of the prepared state without and with the decoherence terms in Eq.~(\ref{eq:Srho}) at $T=3.3\pi$.}
    \label{tabS1}
    \begin{tabular}{lcc} \hline\hline
			Constraints & 
			$\mathbb{F}$ (w/o decoherence) &
			$\mathbb{F}$ (with decoherence) \\ \hline
			$J_{\rm max}=2,\, g_{\rm max}=1$ & 0.8952 & 0.8937\\
			$J_{\rm max}=2,\, g_{\rm max}=2$ & 0.9956 & 0.9939\\
			$J_{\rm max}=2,\, g_{\rm max}=4$ & 0.9994 & 0.9977\\
			 \hline\hline
    \end{tabular}
\end{table}
It can be seen that the fidelity is reduced only by $\sim0.0017$ in all three cases. This result confirms our analysis in the Discussion section of the main paper that the decoherence effects can be neglected in our QOC approach. 

}


\begin{thebibliography}{99}
\bibitem{parameter_unit} The qubit-cavity coupling $g(t)$ and the cavity hopping rate $J(t)$ used in our simulation are in dimensionless units. Because the coupling and hopping strengths in superconducting devices are on the order of a few hundreds of MHz, we assume that $g=1$ corresponds to $2\pi\times100$ MHz when estimating the total evolution time.
\end{thebibliography}
\end{document}